\documentclass[a4paper]{jpconf}
\usepackage{graphicx}
\begin{document}
\title{2D and 3D cubic monocrystalline and polycrystalline materials: their stability and mechanical properties}

\author{C. Jasiukiewicz, T. Paszkiewicz, S. Wolski}

\address{Faculty of Mathematics and Applied Physics,
Rzesz{\'o}w University of Technology, ul. W. Pola 1, PL-35-959 Rzesz{\'o}w, Poland }

\ead{czjas@prz.edu.pl, tapasz@prz.edu.pl, wolan@prz.edu.pl}
%================================================================================================================================================%
\begin{abstract}
We consider 2- and 3-dimensional cubic monocrystalline and polycrystalline materials. Expressions for Young's and shear moduli and Poisson's ratio are expressed in terms of eigenvalues of the stiffness tensor. Such a form is well suited for studying properties of these mechanical characteristics on sides of the stability triangles. For crystalline high-symmetry directions lines of vanishing Poisson's ratio are found. These lines demarcate regions of the stability triangle into areas of various auxeticity properties. The simplest model of polycrystalline 2D and 3D cubic materials is considered. In polycrystalline phases the region of complete auxetics is larger than for monocrystalline materials. 
\end{abstract}
%================================================================================================================================================%
\section{Introduction} 
\label{sc:introd}
In our recent papers we considered mechanical characteristics of 3D crystalline structures of high and middle symmetry \cite{pawol1} and of all 2D symmetry systems \cite{japawol1}. In particular we derived explicit expressions for Young's $E$ and shear $G$ moduli as well as for Poisson's ratio $\nu$ depending on directional cosines of angles between directions $\bf{n}$ of the load and the direction $\bf{m}$ of the lateral strain, respectively, and directions of the crystalline symmetry axes.

Our particular attention was paid to 3D and 2D cubic structures \cite{pawol2}, \cite{japawol2}. All initially unstrained quadratic and cubic materials are characterized by three parameters $s_{1}$, $s_{2}$ and $s_{3}$ belonging to a half-infinite ($s_{1}>0$) prisms with (stability) triangles (ST) in the base lying in the ($s_{2}$,$s_{3}$)-plane. 

Among mechanical characteristics, Poisson's ratio $\nu$ is particularly interesting because the phenomenon of a negative $\nu$, i.e. solids expanding transversely to applied tensile  stress, is counterintuitive. Elastic materials with negative $\nu$ are termed auxetics. Materials with negative $\nu$ for all pairs of vectors $\bf{n}$, $\bf{m}$ are $\it{complete}$ auxetics. Materials are auxetics if there exist pairs of $\bf{n}$, $\bf{m}$ for which $\nu<0$. Materials are non-auxetics if $\nu>0$ for all pairs of vectors $\bf{n}$, $\bf{m}$. In papers \cite{pawol2}, \cite{japawol2} we established regions of ST in which crystalline cubic and quadratic materials are completely auxetic, auxetic and non-auxetic. One may expect that complete auxetics are particularly promising from the standpoint of technology.

Many technically important materials are polycrystalline. In view of potential applications of crystalline auxetic materials, the question how polycrystallinity influences the auxetic properties deserves some attention. 

The elastic properties of polycrystalline materials depend on the single-crystal stiffnesses (elastic constants) of the crystallites which build up the polycrystal and on the manner in which crystallites are connected. In most cases exact orientations, shapes and connections are not known. These gaps in knowledge can be overcome  with the help of orientation and grain shape distribution functions. Generally, one has to resort to more or less realistic assumptions (cf. \cite{hirsekorn}). We assume that the grains have the same shape and volume and that their orientations are completely random.  
%%%%%%%%%%%%%%%%%%%%%%%%%%%%%%%%%%%%%%%%%%%%%%%%%%%%%%%%%%%%%%%%%%%%%%%%%%%%%%%%%%%%%%%%%%%%%%%%%%%%%%%%
%================================================================================================================================================%
\section{The stability conditions for initially unstressed crystalline quadratic and cubic media} 
\label{sect:stab-cond} 
%================================================================================================================================================%
Mechanical properties of crystalline materials depend on components of the stiffness tensor $\bf{C}$ or on the components of the compliance tensor $\bf{S}$. These tensors are mutually inverse, i.e. 
\begin{equation}
	\bf{C}\bf{S}=\bf{S}\bf{C}={\bf{I}}_{4},
\label{eq:inverse}
\end{equation}
where $\left({\bf{I}}^{(4)}_{ij,kl}\right)=\left(\delta_{ik}\delta_{jl}+\delta_{ij}\delta_{jk}\right)/2$. The product of two fourth rank tensors $A$ and $B$ has components $({\bf{AB}})_{ijkl}=\sum_{r,s=1}^{3}A_{ij,rs}B_{rs,kl}$. 

In this paper we consider only 2D and 3D cubic crystalline or polycrystalline materials. For mechanically stable initially unstrained materials both $\bf{C}$ and $\bf{S}$ tensors are positive, therefore their eigenvalues $c_{I}$, $s_{I}$ $(I=J,L,M)$ are positive too (\cite{pawol1}, \cite{japawol1}, \cite{walpole})
\begin{equation}
	c_{I}>0, s_{I}>0\; (I=J,M,L).
	\label{eq:positive}
\end{equation}
Because of Eq. (\ref{eq:inverse}) $s_{I}=c^{-1}_{I}$ vanishing eigenvalues $c_{I}\; (I=J,L,M)$ signal instabilities related to phase transitions \cite{papruziel}.

In the case of quadratic materials \cite{japawol1} 
\begin{equation}
	c_{J}=C_{11}+C_{12}, \; c_{L}=C_{11}-C_{12}, c_{M}=2C_{66}.
\label{eq:2d-eigenv}
\end{equation}
In the case of cubic materials \cite{pawol1}, \cite{walpole}
\begin{equation}
	c_{J}=C_{11}+2C_{12}, \; c_{L}=C_{11}-C_{12}, c_{M}=2C_{66}.
\label{eq:3d-eigenv}
\end{equation}
In Eqs. (\ref{eq:2d-eigenv}) and (\ref{eq:3d-eigenv}) we used the familiar Voigt's notation \cite{nye}: $C_{11}=C_{11,11}$, $C_{12}=C_{11,22}$ and $C_{66}=C_{12,12}$. 

For cubic materials a more reasonable choice of independent parameters was proposed by Every \cite{every} and for quadratic materials in our paper \cite{japawol2}
\begin{eqnarray}
	s_{1}=C_{11}+(d-1)C_{66},\, s_{2}=\frac{C_{11}-C_{66}}{s_{1}},\nonumber \\ 
	 s_{3}=\frac{C_{11}-C_{12}-2C_{66}}{s_{1}},
	\label{eq:every}
\end{eqnarray}
where $d=2$ for 2D, $d=3$ for 3D materials. 

\begin{figure}[htb]
	\centering
		\includegraphics[width=1.00\textwidth]{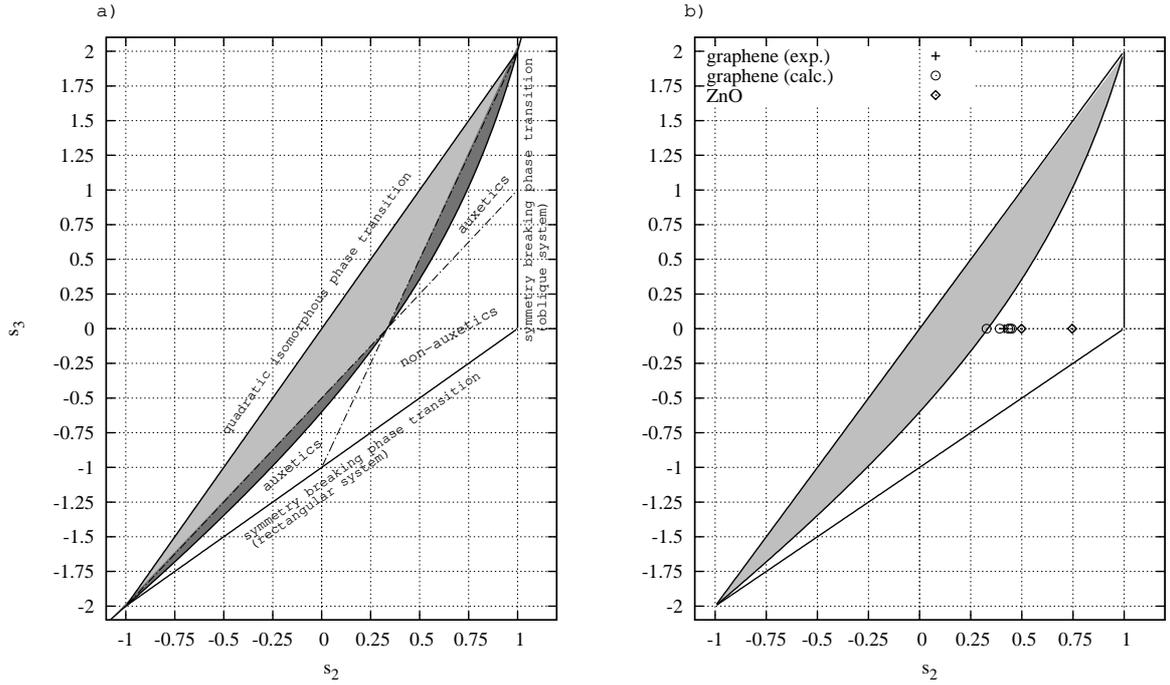}
	\caption{a) Division of the stability triangle for quadratic materials into areas of various auxetic properties. b)  The shaded area is the region of complete polycrystaline auxetics. Locations of results for graphene~\cite{falkovsky},~\cite{michel_pr} and graphene-like ZnO~\cite{tu} are indicated.}
	\label{fig:triangles}
\end{figure}

We rewrite expressions (\ref{eq:2d-eigenv}) and (\ref{eq:3d-eigenv}) using Every's parameters $s_{1}$, $s_{2}$ and $s_{3}$. In the case of quadratic materials we get
\begin{equation}
c_{J}=s_{1}\left(2s_{2}-s_{3}\right),\; c_{L}=s_{1}\left(1-s_{2}+s_{3}\right),\; s_{M}=2s_{1}\left(1- s_{2}\right),
	\label{eq:2d-eigenval-every}
\end{equation}
whereas in the case of cubic materials 
\begin{equation}
	c_{J}=\frac{s_{1}}{3}\left(10s_{2}-6s_{3}-1\right),\; c_{L}=2\frac{s_{1}}{3}\left(1-s_{2}\right),\; s_{M}=2\frac{s_{1}}{3}\left(3s_{3}- 2s_{2} +2\right).
	\label{eq:3d-eigenval-every}
\end{equation}
For both cubic and quadratic materials the first of the inequalities (\ref{eq:positive}) define stability regions in the form of semi-infinite prisms ($s_{1}> 0$) with two distinct triangles (STs) at the base in $s_{2}$, $s_{3}$-plane.
\begin{equation}
	s_{1}>0, \, s_{2}<1,
	\label{eq:every_inequal}
\end{equation}
\begin{eqnarray}
	2D: \left(1-s_{2}+s_{3}\right)>0,\, \left(s_{2}-s_{3}/2\right)>0, \nonumber \\
	3D: \left(1 + 2s_{2}\right)>\left|4s_{2}-3s_{3}-1\right|,\, \left(10s_{2}-6s_{3}\right)>1.
	\label{eq:every_inequal_remaining}
\end{eqnarray}
Both stability triangles are lying in the $(s_{2},s_{3})$-plane (Fig. \ref{fig:triangles}).
The sides of the stability triangles (STs) are defined by lines
\begin{eqnarray}
	2D:\;\;\; s^{(2D,1)}_{3}=2s_{2},\; s^{(2D,2)}_{3}=s_{2}-1,\; s^{(2D,3)}_{2}=1,\nonumber \\
	3D:\;\;\; s^{(3D,1)}_{3}=5s_{2}/3-1/6, \; s^{(3D,2)}_{3}=2s_{2}/3-2/3
	,\; s^{(3d,2)}_{2}=1.
\label{eq:triangle-lines}	
\end{eqnarray}

\begin{figure}[htpb]
\centering
		\includegraphics[width=1.00\textwidth]{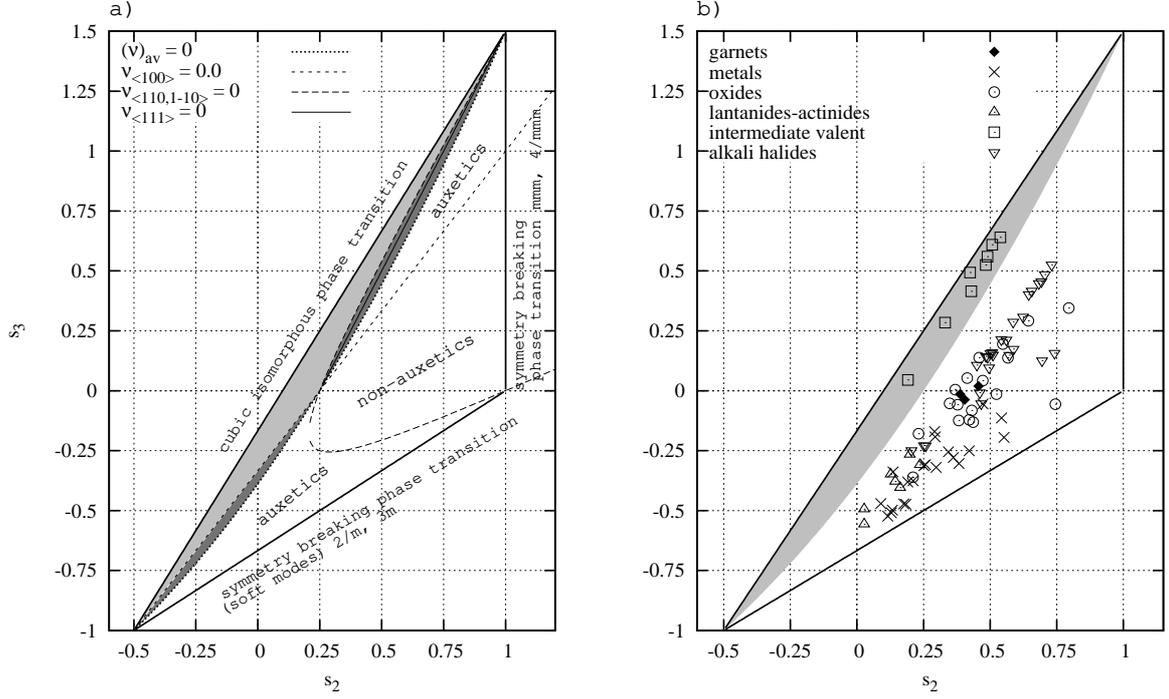}
 
	\caption{a) Division of the stability triangle for cubic materials into areas of various auxetic properties. b)  The shaded area is the region of complete polycrystaline auxetics. Locations of several classes of elastic materials are indicated (cf. ~\cite{pawol3}).}
	\label{fig:poly_triangle}
\end{figure}

A point of the stability triangle together with the value of $s_{1}$ represents an existing or potentially existing crystalline material.   
 For acoustic and elastic characteristics the parameter $s_{1}$ is a scaling parameter (cf. \cite{papruziel}, \cite{papru}). In Figs.\ref{fig:triangles} and \ref{fig:poly_triangle} we show changes of the division of STs in two regions into various auxetic properties. Points of area with lighter shade  represent complete monocrystaline auxetics, whereas in the area with darker shade lie complete polycrystaline auxetics. 

For symmetry systems other than cubic and isotropic, as well as for the oblique symmetry system, elastic properties are defined in spaces of dimensions higher than 3. In the case of rectangular symmetry system the space of parameters can be effectively reduced to a 3D space (cf. Appendix). Monocrystalline stable {\it isotropic} materials are located on the interval of the lines defined by equation $s_{3}=0$ belonging to the appropriate triangle.
%================================================================================================================================================%
\section{Mechanical characteristics of quadratic and cubic materials}
\label{sect:mechan_charct_monocr}  
%================================================================================================================================================%
Mechanical properties of an elastic body are characterized by Young's and shear moduli and Poisson's ratio. Recent interest in crystalline materials has made the understanding these characteristics increasingly important. In papers \cite{pawol1} and \cite{japawol1} we derived expressions for them which do not depend on the choice of Cartesian coordinate system. The Young modulus $E({\bf n})$ depends on the (direction) cosines that unit vector ${\bf n}$ of the load makes with the crystalline axes. The shear modulus $G$, as well as the Poisson's ratio $\nu)$, additionally depend on the direction cosines of the vector of lateral load ${\bf m}$.

Expressions for these characteristics can be written in terms of the eigenvalues $c_{J}$, $c_{L}$ and $c_{M}$
For quadratic materials we obtain

\begin{equation}
	E_{q}({\bf n})=\frac{2c_{J}c_{L}c_{M}}{c_{L}c_{M}+c_{J}c_{M}\left(n_{1}^{2}-n_{2}^{2}\right)^{2}+4c_{J}c_{L}n_{1}^{2}n_{2}^{2}},
\label{eq:E_quadratic}
\end{equation}
\begin{equation}
	G_{q}({\bf n},{\bf m})=\frac{1}{2}\frac{c_{L}c_{M}}{c_{M}\left(m_{1}n_{1}-m_{2}n_{2}\right)^{2}+c_{L}\left(m_{1}n_{2}+m_{2}n_{1}\right)^{2}},
\label{eq:G_quadratic}
\end{equation}
\begin{equation}
-\nu_{q}({\bf n},{\bf m})=\frac{c_{L}c_{M}+c_{J}c_{M}\left(m_{1}^{2}-m_{2}^{2}\right)\left(n_{1}^{2}- n_{2}^{2}\right)+4c_{J}c_{L}m_{1}m_{2}n_{1}n_{2}}{c_{L}c_{M}+c_{J}c_{M}\left(n_{1}^{2}- n_{2}^{2}\right)^{2}+4c_{J}c_{L}n_{1}^{2}n_{2}^{2}},
\label{eq:nu_quadratic}
\end{equation}

For cubic materials we get
\begin{equation}E_{c}({\bf n})=\frac{3c_{J}c_{L}c_{M}}{\left(c_{L}c_{M}-c_{J}c_{L}+3c_{J}c_{M}\right)+3c_{J}\left(c_{L}-c_{M}\right)\sum_{i=1}^{3}n_{i}^{4}},
\label{eq:E_cubic}
\end{equation}
\begin{equation}
G_{c}({\bf n},{\bf m})=\frac{c_{J}c_{L}c_{M}}{c_{J}c_{M}+2c_{J}\left(c_{L}-c_{m}\right)\sum_{i=1}^{3}m_{i}^{2}n_{i}^{2}},
\label{eq:G_cubic}
\end{equation}
\begin{equation}
-\nu_{c}({\bf n},{\bf m})=\frac{c_{L}\left(c_{M}-c_{J}\right)+3c_{J}\left(c_{L}-c_{M}\right)\sum_{i=1}^{3}m_{i}^{2}n_{i}^{2}}{\left(c_{L}c_{M}-c_{J}c_{L}+3c_{J}c_{M}\right)+ 3c_{J}\left(c_{L}-c_{M}\right)\sum_{i=1}^{3}n_{i}^{4}}.
\label{eq:nu_cubic}	
\end{equation}

Note that expressions (\ref{eq:E_quadratic}-\ref{eq:nu_cubic}) allow one to calculate $E$, $G$ and $\nu$ for a given pair of direction vectors $({\bf m},{\bf n})$ and for chosen values of $s_{1}$, $s_{2}$ and $s_{3}$. In the case of polycrystalline materials numerical calculation provides one with analogous results. Performing calculations for a mesh of points of STs we obtained the appropriate maps \cite{japawol3}. These maps were calculated for $s_{1}=1$. 

The direction vector of applied tension $\bf{n}$ and direction vector $\bf{m}$ in which the lateral expansion/contraction is measured are mutually perpendicular, i.e. ${\bf{nm}}=0$. For quadratic materials 
\begin{eqnarray}
\label{eq:m_vector1}
n_{1}=m_{2}=\cos\varphi,\nonumber\;\\  
n_{2}=-m_{1}=\sin{\varphi},
\label{eq:quadr_angle_inequalities}
\end{eqnarray}
where $0<\varphi\leq 2\pi$.

In the case of cubic materials \cite{aouni_wheeler}
\begin{eqnarray}
\label{eq:n_vector}
n_{1}=\cos{\alpha}\cos{\varphi}\cos{\theta}-\sin{\alpha}\sin{\theta},\nonumber \;\\  
n_{2}=\cos{\alpha}\cos{\varphi}\sin{\theta}+\sin{\alpha}\cos{\theta},\nonumber \;\\ 
n_{3}=-\cos{\alpha}\sin{\varphi},
\label{eq:cubic_n}
\end{eqnarray}
and 
\begin{eqnarray}
\label{eq:m_vector}
m_{1}=\sin{\varphi}\cos{\theta},\nonumber\;\\  
m_{2}=\sin{\varphi}\sin{\theta},\nonumber\;\\ 
m_{3}=\cos{\varphi},
\label{eq:cubic_m}
\end{eqnarray}
where
\begin{equation}
 -\pi<\alpha\leq\pi,\, 0<\varphi\leq\pi,\, 0<\theta\leq\ 2\pi.
\label{eq:cubic_angle_inequalities}
\end{equation}
%================================================================================================================================================%
\section{Mechanical characteristics on the sides of stability triangles}
\label{sect:mech_char_on_sides}
%================================================================================================================================================%
For directions of high symmetry the mechanical characteristics can vanish, signaling phase transitions (cf. \cite{papruziel}). New phases are indicated in Fig. \ref{fig:triangles}.

Consider quadratic materials. For ${\bf n}^{(q)}_{1}=\left\langle 1 1\right\rangle$, ${\bf m}^{(q)}_{1}=\left\langle 1 \bar{1}\right\rangle$ 
\begin{equation}
	E_{q}=\frac{2c_{J}c_{M}}{c_{J}+c_{M}},\; G_{q},=\frac{c_{L}}{2},\; \nu_{q}=\frac{c_{J}-c_{M}}{c_{J}+c_{M}}.
	\label{quadr_n1}
\end{equation}
This means that for $E_{q}({\bf n}_{1},{\bf m}_{1})$ vanishes on the top and on the vertical side of ST, whereas $G_{q}({\bf n}_{1},{\bf m}_{1})$ vanishes on the bottom side of it. 

For ${\bf n}_{2}=\left\langle 1\, 0\right\rangle$, ${\bf m_{2}}=\left\langle 0\; 1 \right\rangle$ 
\begin{equation}
	E_{q}=\frac{2c_{J}c_{L}}{c_{J}+c_{L}},\; G_{q}=\frac{c_{M}}{2},\; \nu_{q}=\frac{c_{J}-c_{L}}{c_{J}+c_{L}}.
	\label{quadr_n2}
\end{equation}

For ${\bf n}_{2}$ and ${\bf m}_{2}$ $E_{q}$ vanishes on the top and bottom side of ST, whereas $G_{q}$ vanishes on the vertical side. 

For ${\bf n}^{(q)}_{3}=\left\langle n_{3}^{(q,1)}\; n_{3}^{(q,2)}\right\rangle$ and ${\bf m}^{(q)}_{3}=\left\langle \bar{n}^{(q,2)}_{3}\; n_{3}^{(q,1)}\right\rangle$ 
\begin{equation}
	E_{q}=\frac{4c_{J}c_{L}c_{M}}{2c_{L}c_{M}+c_{J}c_{L}+c_{J}c_{M}},\;
	G_{q}=\frac{c_{L}c_{M}}{c_{L}+c_{M}},\;\; \nu_{q}=\frac{c_{J}\left(c_{L}+c_{M}\right)-2c_{L}c_{M}}{c_{J}\left(c_{L}+c_{M}\right)+2c_{L}c_{M}},
	\label{eq:quadr_n3}
\end{equation}
where $n_{3}^{(q,1)}=\cos(\pi/8)$ and $n_{3}^{(q,2)}=\sin(\pi/8)$. This means that $E_{q}({\bf n}_{3}^{(q)})$ vanishes on all sides of ST, whereas $G_{q}({\bf n}_{3}^{(q)},{\bf m}_{3}^{(q)})$ vanishes on the bottom and vertical side of ST. 

Similar results are valid for cubic materials. For ${\bf n}^{(c)}_{1}=\left\langle 0\; 0\; 1\right\rangle,\,{\bf m}^{(c)}_{1}=\left\langle 1 \; 1 \; 0\right\rangle$ Eqs. (\ref{eq:E_cubic}), (\ref{eq:G_cubic}) and (\ref{eq:nu_cubic}) give 
\begin{equation}
{\bf n}^{(c)}_{1},\,{\bf m}^{(c)}_{1}:\;\; E_{c}=\frac{3c_{J}c_{M}}{2c_{L}+c_{M}},\; G_{c}=\frac{c_{L}}{2},\; \nu_{c}=\frac{c_{J}-c_{M}}{2c_{J}+c_{M}}, 
\label{eq:cubic_n1}	
\end{equation}
hence $E_{c}({\bf n}^{(c)}_{1})=0$ on the top and bottom side of ST, $G_{c}({\bf n}^{(c)}_{1}, {\bf m}^{(c)}_{1})=0$ on the vertical side of it. For ${\bf n}_{2}^{(c)}=\left\langle 1\,1\,0\right\rangle$ , ${\bf m}_{2}^{(c)}=\left\langle \bar{1}\,1\,0\right\rangle$ we obtain
\begin{eqnarray}
{\bf n}^{(c)}_{2}\;\;{\bf m}^{(c)}_{2}:\;\; E_{c}=\frac{3c_{J}c_{L}c_{M}}{c_{J}c_{L}/2+3c_{J}c_{M}/2+c_{L}c_{M}/2}, \nonumber \\
G_{c}=\frac{c_{M}}{2},\;\;  -\nu_{c}=\frac{c_{J}c_{L}/2-3c_{J}c_{M}/2+c_{L}c_{M}}{c_{J}c_{L}/2+3c_{J}c_{M}/2+c_{L}c_{M}}. 
\label{eq:cubic_n2}	
\end{eqnarray}  
For ${\bf n}^{(c)}_{2}$ and ${\bf m}^{(c)}_{2}$ Young's modulus vanishes on all sides of ST, whereas the shear modulus vanishes on the bottom side of it. Finally, for ${\bf n}^{(c)}_{3}=\left\langle 1\; 1\; 1\right\rangle,\,{\bf m}^{(c)}_{3}=\left\langle 0 \; 1 \; \bar{1}\right\rangle$
\begin{equation}
{\bf n}^{(c)}_{3},\;{\bf m}^{(c)}_{3}:\;\; E_{c}=\frac{3c_{J}c_{L}}{2c_{J}+c_{L}}, 
G_{c}=\frac{3c_{L}c_{M}}{2c_{L}+c_{M}},\;\;  \nu_{c}=\frac{c_{J}-c_{L}}{c_{L}-2c_{J}}. 
\label{eq:cubic_n3}	
\end{equation}
For ${\bf n}^{(c)}_{3}$ and ${\bf m}^{(c)}_{3}$ Young's modulus vanishes on the top and vertical side of ST, the shear modulus vanishes on the vertical and bottom side. 

For all considered high symmetry directions Young's and shear moduli are non-negative. 

%================================================================================================================================================%
\section{Lines of vanishing Poisson's ratio}
\label{sect:lines_vanishing_nu}
%================================================================================================================================================%
For a chosen direction ${\bf n}$ of load and lateral strain ${\bf m}$ the general expressions presented in Sect. \ref{sect:mechan_charct_monocr} allow us to find the lines of vanishing $\nu$ that divide the stability triangles into regions of various auxeticity properties. These lines are particularly interesting for ${\bf n}$ and ${\bf m}$ directed along high symmetry directions. 

Consider the quadratic materials. For the following pairs (${\bf n}$, ${\bf m}$) the condition $\nu_{q}\left({\bf m},{\bf n}\right)=0$ leads to equations 
\begin{eqnarray}
\left({\bf n}^{(q)}_{1},\;{\bf m}^{(q)}_{1}\right):\; s_{3}(s_{2})=2\left(s_{2}-1\right),\nonumber \\
\left({\bf n}^{(q)}_{2},\;{\bf m}^{(q)}_{2}\right):\; s_{3}(s_{2})=3s_{2}/2-1/2. 	
\label{eq:quadr_vanishing_nu_lines}
\end{eqnarray}
For cubic materials we get 
\begin{eqnarray}
\left({\bf n}^{(c)}_{1}, \; {\bf m}^{(c)}_{1}\right): \;s_{3}\left(s_{2}\right)=4s_{2}/3-1,\nonumber\\
\left({\bf n}^{(c)}_{2}, \; {\bf m}^{(c)}_{2}\right): \;s_{3}\left(s_{2}\right)=\frac{7}{6}s_{2}-\frac{5}{12}\pm \frac{1}{12}\sqrt{68s^{2}_{2}+20s_{2}-7},\nonumber\\
\left({\bf n}^{(c)}_{3}, \; {\bf m}^{(c)}_{3}\right): \;s_{3}\left(s_{2}\right)=2s_{2}-1.
\label{eq:cubic_vanishing_nu_lines}	
\end{eqnarray}
All the above lines of vanishing $\nu$ are shown in Fig. \ref{fig:triangles}.

In our papers \cite{pawol2}, \cite{japawol2} we introduced the division of the stability triangle into regions of complete auxeticity, auxeticity and non-auxeticity. The above regions are indicated on diagrams in Fig. \ref{fig:triangles}. Inspecting them one can distinguish lines of vanishing $\nu$ studied in this section.  
%%%%%%%%%%%%%%%%%%%%%%%%%%%%%%%%%%%%%%%%%%%%%%%%%%%%%%%%%%%%%%%%%%%%%%%%%%%%%%%%%%%%%%%%%%%%%%%%%%%%%%%%%%%%%%%%%%%%%%%%%%%%%%%%%%%%%%%%%%%%%%%%%%%%%	
\section*{Mechanical characteristics of 2- and 3-d cubic polycrystalline materials}
\label{sc:poly}
%%%%%%%%%%%%%%%%%%%%%%%%%%%%%%%%%%%%%%%%%%%%%%%%%%%%%%%%%%%%%%%%%%%%%%%%%%%%%%%%%%%%%%%%%%%%%%%%%%%%%%%%%%%%%%%%%%%%%%%%%%%%%%%%%%%%%%%%%%%%%%%%%%%%
Following the assumptions made at the end of Sect. \ref{sc:introd} to find the mechanical characteristics of polycrystalline materials one should calculate mean values of expressions defining them. Consider quadratic materials and a function $F$ of the angle $\varphi$, then in agreement with definition (\ref{eq:m_vector1})
\begin{equation}
	\left\langle F\right\rangle=\frac{1}{2\pi}\int_{0}^{2\pi}d\varphi F(\varphi).
\label{eq:quadr_average}
\end{equation}

Calculating $\left\langle E_{q}\right\rangle$, $\left\langle G_{q}\right\rangle$ and $\left\langle \nu_{q}\right\rangle$ we obtained 
\begin{equation}
\left\langle E_{q}\right\rangle=\frac{4c_{J}c_{L}c_{M}}{\sqrt{\left[2c_{L}c_{M}+c_{J}\left(c_{L}+c_{M}\right)\right]^{2}-c_{J}^{2}\left(c_{L}-c_{M}\right)^{2}}},
	\label{eq:E-q-av}
\end{equation}
\begin{equation}
	\left\langle G_{q}\right\rangle=\sqrt{c_{L}c_{M}}/2,
	\label{eq:G-q-av}
\end{equation}
\begin{equation}
	\left\langle \nu_{q}\right\rangle=1-\frac{4c_{L}c_{M}}{\sqrt{\left[2c_{L}c_{M}+c_{J}\left(c_{L}+c_{M}\right)\right]^{2}-c_{J}^{2}\left(c_{L}-c_{M}\right)^{2}}}.
	\label{eq:nu-q-av}
\end{equation}

Table \ref{table} summarizes the properties of mean values of the mechanical characteristics and eigenvalues $c_{I}\; \left(I=J,L,M\right)$ on sides of ST. These properties align with calculated maps of $\left\langle E_{q}\right\rangle$, $\left\langle G_{q}\right\rangle$ and $\left\langle \nu_{q}\right\rangle$ \cite{japawol3}.
\begin{table}[h]
\caption{Values of $\left\langle E_{q}\right\rangle$, $\left\langle G_{q}\right\rangle$ and $\left\langle \nu_{q}\right\rangle$ on the sides of ST. J -- top side, M -- bottom side, L -- vertical side; $U=J,L,M$.}
\begin{center}
\begin{tabular}
[c]{l|l|l|l|l}%
Side  & $c_{U}$      & $\left\langle E_{q}\right\rangle$ & $\left\langle G_{q}\right\rangle$ & $\left\langle \nu_{q}\right\rangle$\\ \hline
J     &0             & 0                                 & $\sqrt{c_{M}c_{L}}/2$             & -1\\ \hline
L     &0             & 0                                 & 0                                 &  \;1\\ \hline
M     &0             & 0                                 & 0                                 &  \;1
\label{table}
\end{tabular}
\end{center}
\end{table}

In the case of quadratic polycrystalline materials the condition $\left\langle \nu_{q}\right\rangle=0$ leads to equations of line 
\begin{equation}
	s_{3}\left( s_{2}\right)=\frac{3s^{2}_{2}-10s_{2}+3}{3s_{2}-5}, 
\label{quadratic_nu_aver_vanishing_line}	
\end{equation}
demarcating the stability triangle to the regions of {\it complete} auxetics and non-auxetics. In case of cubic materials such a line is found numerically. Inspecting Fig. \ref{fig:poly_triangle} we note that for polycrystalline 2D and 3D cubic materials the area of region of complete auxeticity is larger than for monocrystalline materials. 

In the case of cubic materials and a function $F$ of angles $\alpha$, $\varphi$ and $\theta$ (cf. Eqs. (\ref{eq:cubic_n}) and (\ref{eq:cubic_m})) the mean value $\left\langle F\right\rangle$ of $F(\alpha,\varphi,\theta)$  is defined as
\begin{equation}
\left\langle F\right\rangle=\frac{1}{8\pi^{2}}\int_{-\pi}^{+\pi}d\alpha\int_{0}^{\pi}{d\varphi}\sin\varphi\int_{0}^{2\pi}d\theta F(\alpha, \varphi, \theta).
\label{eq:cub_average}
\end{equation}

Since we consider the model of randomly oriented crystallites, the averaged mechanical characteristics are isotropic. This means, that if $\left\langle \nu\right\rangle>0$, one deals with isotropic non-auxetics, whereas if $\left\langle \nu\right\rangle<0$, a polycrystal is a complete isotropic auxetic. We conclude that in the case of considered models of polycrystals the division of the stability triangles simplifies.
%===============================================================================================================================================%
\section{Concluding remarks}
\label{sect:conclusions}
We have shown that mechanical characteristics of elastic materials expressed in terms of eigenvalues of the stiffness tensor are well suited for the study of their general properties. The simplest model of polycrystalline 2D and 3D cubic materials indicates that some monocrystalline auxetics become  complete auxetics in the polycrystalline phase.   
%===============================================================================================================================================%
%%%%%%%%%%%%%%%%%%%%%%%%%%%%%%%%%%%%%%%%%%%%%%%%%%%%%%%%%%%%%%%%%%%%%%%%%%%%%%%%%%%%%%%%%%%%%%%%%%%%%%%%%%%%%%%%%%%%%%%%%%%%%%%%%%%%%%%%%%%%%%%%%%%%%	
\section*{Appendix}
\label{sc:appendix}
%%%%%%%%%%%%%%%%%%%%%%%%%%%%%%%%%%%%%%%%%%%%%%%%%%%%%%%%%%%%%%%%%%%%%%%%%%%%%%%%%%%%%%%%%%%%%%%%%%%%%%%%%%%%%%%%%%%%%%%%%%%%%%%%%%%%%%%%%%%%%%%%%%%%
Elastic properties of rectangular materials are characterized by four stiffnesses $C_{11}$, $C_{22}$, $C_{12}$, $C_{66}$ (cf. \cite{japawol1} and \cite{musgrave}). Initially unstrained rectangular materials are stable if 
\begin{equation}
	C_{11}>0,\;C_{22}>0,\;C_{66}>0,\;C_{11}C_{22}>C_{12}^{2}.
	\label{eq:rectan_stab}
\end{equation}
Introduce four new parameters
\begin{equation}
	s_{1}=\left(C_{11}+C_{22}\right)/2+C_{66},
	\label{eq:rectan_s1}
\end{equation}
\begin{equation}
	s_{2}=\left[\left(C_{11}+C_{22}\right)/2-C_{66}\right]/s_{1},
	\label{eq:rectan_s2}
\end{equation}
\begin{equation}
	s_{3}=\left[\left(C_{11}+C_{22}\right)/2-C_{12}-2C_{66}\right]/s_{1},
	\label{eq:rectan_s3}
\end{equation}
\begin{equation}
	s_{4}=\left(C_{11}-C_{12}\right)/2s_{1}.
	\label{eq:rectan_s4}
\end{equation}
Because of inequalities (\ref{eq:rectan_s1})-(\ref{eq:rectan_s4})
\begin{equation}
	s_{1}>0,\; \; -1<s_{2}<1,
	\label{eq:s_inequ} 
\end{equation}
$s_{4}$ can be positive and negative. 

\begin{figure}[htpb]
\centering
		\includegraphics[width=0.80\textwidth]{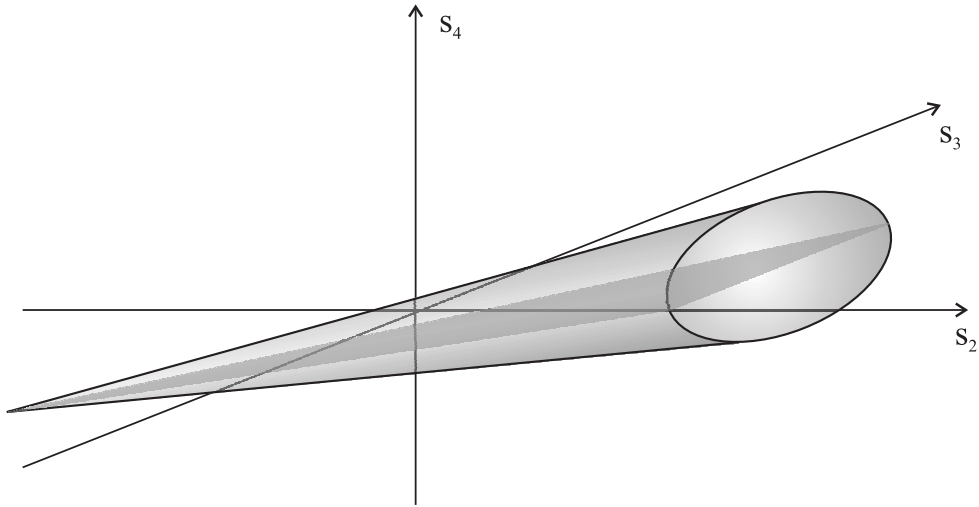} 
	\caption{Rectangular materials: mechanical stability region in $\left(s_{2},s_{3},s_{4}\right)$-space.}
	\label{fig:cone}
\end{figure} 

Consider 3D space of parameters $s_{2}$, $s_{3}$, $s_{4}$. In planes $s_{2}=\alpha$ ($-1<\alpha<1$) the last of inequalities (\ref{eq:rectan_stab}) takes the form 
\begin{equation}
	s_{4}^{2}+\left(s_{3}-d_{1}\right)^{2}<d_{2}^{2},
	\label{eq:rect_ineq_3D}
\end{equation}
where $d_{1}\left(\alpha\right)=\left(3\alpha-1\right)/2$, $d_{2}\left(\alpha\right)=\left(1+\alpha\right)/2$. Inequality (\ref{eq:rect_ineq_3D}) defines a section of the oblique cone with the apex located at the point $\left(s_{2}=-1,s_{3}=-2, s_{4}=0\right)$. For $-1<\alpha<1$ the cross-sections of the cone perpendicular to $s_{2}$ axis are circles with radii equal to $d_{2}(\alpha)$ and centers $\left(s_{2}=\alpha,\; s_{3}=d_{1}(\alpha)\right),\; s_{4}=0$ (Fig. \ref{fig:cone}). For $s_{4}=0$ the cone (\ref{eq:rect_ineq_3D}) reduces to ST defined by Eq. (\ref{eq:every_inequal_remaining}). 
%===============================================================================================================================================%
\section*{Conclusions}
%===============================================================================================================================================%

\end{document}